\documentclass{svjour3}  
\smartqed  % flush right qed marks, e.g. at end of proof

\usepackage{amsmath}
\usepackage{graphicx}
\usepackage{color}
\usepackage{graphicx}

\usepackage[colorlinks=true,linkcolor=blue,citecolor=blue]{hyperref}

\newcommand{\be}{\begin{equation}}
\newcommand{\ee}{\end{equation}}
\newcommand{\bea}{\begin{eqnarray}}
\newcommand{\eea}{\end{eqnarray}}

\newcommand{\eref}[1]{Eq.~(\ref{#1})}%
\newcommand{\fref}[1]{Fig.~\ref{#1}} %
\newcommand{\sref}[1]{Sec.~\ref{#1}}%
\newcommand{\tref}[1]{Table~\ref{#1}}%

\begin{document}

\title{Shock propagation following an intense explosion in an inhomogeneous gas: core scaling and hydrodynamics}      
\titlerunning{Blast waves}
\author{Amit Kumar \and R. Rajesh}

\institute{Amit Kumar \at
              	The Institute of Mathematical Sciences, CIT Campus, Taramani, Chennai 600113, India \\
              	Homi Bhabha National Institute, Training School Complex, Anushakti Nagar, Mumbai 400094, India\\
               	\email{kamit@imsc.res.in}  	    
         \and
             R. Rajesh \at
              	The Institute of Mathematical Sciences, CIT Campus, Taramani, Chennai 600113, India \\
              	Homi Bhabha National Institute, Training School Complex, Anushakti Nagar, Mumbai 400094, India\\
               	\email{rrajesh@imsc.res.in}           
}

\date{Received: \today / Accepted: }

\maketitle

\begin{abstract}
We study the shock propagation in a spatially inhomogeneous gas following an intense explosion. We generalize the exact solution of the Euler equation for the spatio-temporal variation of density, velocity, and temperature to arbitrary dimensions. From the asymptotic behavior of the solution near the shock center, we argue that only for a critical dimension dependent initial density distribution will the Euler equation provide a correct description of the problem. For general initial density distributions, we use event-driven molecular dynamics simulations in one dimension to demonstrate that the Euler equation fails to capture the behavior near the shock center. However, the Navier-Stokes equation successfully resolves this issue. The crossover length scale below which the dissipation terms are relevant and the core scaling for the data near the shock center are derived and confirmed in EDMD simulations.
\end{abstract}

\section{\label{sec1-Introduction}Introduction}
The exact solution of Euler equation to study the evolution of thermodynamic quantities caused by an intense explosion is a classic problem in gas dynamics~\cite{landaubook,barenblatt1996scaling,whitham2011linear}. After an initial radiating regime, the systems cools down to reach a hydrodynamic regime where the transport of energy takes place mainly due to the motion of the particles. The sudden release of energy in a localized region results in a spherically symmetric disturbance growing in time with a shock front separating the moving  gas from the ambient gas. Across the shock front density, velocity, temperature, and pressure are discontinuous~\cite{landaubook,whitham2011linear,sedov_book}. The evolution of these quantities has been studied both for a homogeneous gas in which initial density of the gas is uniform, as well as an inhomogeneous gas where the initial density, $\rho(r)$,  at distance $r$ from the  explosion varies  as $\rho(r)=\rho_0 r^{-\beta}$. 

In the scaling regime, from dimensional analysis, it is straightforward to obtain that the radius of the shock front, $R(t)$, grows as  $R(t)\sim (E_0/\rho_0)^{\alpha/2}t^{\alpha}$ with $\alpha=2/(2+d-\beta)$ in $d$-dimensions, where $E_0$ is the energy input~\cite{sedov_book,taylor1950formation,taylor1950formation2,jvneumann1963cw,sedov1946}. The power law growth of $R(t)$ in a homogeneous medium ($\beta=0$) has been verified in the trinity explosion~\cite{taylor1950formation,taylor1950formation2}, and in blast waves due to the high energy laser pulses in gas jets in $d=2$~\cite{edwards2001investigation}, in plasma media in $d=3$~\cite{edens2004study}, and in the cluster of different gases in $d=2,3$~\cite{moore2005tailored}.

The spatio-temporal evolution of the thermodynamic quantities are studied using the continuity equations of mass, momentum, and the energy, corresponding to the  Navier-Stokes equation. However, in the scaling limit $r\to \infty$, $t\to \infty$ keeping $r t^{-\alpha}$ finite, the heat conduction and viscosity terms are negligible, leading to the Euler equation. If the gas is assumed to an ideal gas, then an exact solution for density, velocity and temperature, can be found both for the homogeneous case~\cite{taylor1950formation,taylor1950formation2,jvneumann1963cw,sedov_book,sedov1946} as well as the inhomogeneous case~\cite{sedov_book}. These solutions have found applications in modeling the supernova explosion and the early stage evolution of young supernova remnants~\cite{gull1973numerical,cowie1977early,bertschinger1983cosmological,bertschinger1985cosmological,cioffi1988dynamics,ostriker1988astrophysical}. The solution can be generalized to that for a continuous source~\cite{dokuchaev2002self,kumar2024shock}, and finds application in the nebulae formation and its motion due to the continuous injection of stellar wind into the interstellar gas~\cite{avedisova1972formation,falle1975numerical}.

There has been a recent renewed interest in the explosion problem in the form of large scale simulations of hard sphere systems in three~\cite{jabeen2010universal,joy2021shock3d,kumar2022blast}, two~\cite{jabeen2010universal,joy2021shock,barbier2016microscopic,kumar2022blast} and one dimensions~\cite{ganapa2021blast,chakraborti2021blast}. Significant differences were found between the exact solution and simulation data near the shock center~\cite{joy2021shock,joy2021shock3d,kumar2022blast}, which was argued to be due to the non-commutation of the limits  -- first taking the scaling limit and then finding the solution, or finding the solution of the Navier-Stokes equation and then taking the scaling limit. It was shown that when dissipation terms are included in the Euler equation giving rise to the Navier-Stokes equation, then the discrepancies of the theory with simulations of an explosion in a homogeneous gas can be accounted for both in one dimension~\cite{ganapa2021blast,chakraborti2021blast} as well as higher dimensions~\cite{kumar2022blast}. A similar resolution has been found in the case of continuously driven shocks~\cite{kumar2024shock}. The crossover behavior of the scaling functions from the Euler solution to the Navier-Stokes solution near the shock center has been quantified in one~\cite{ganapa2021blast,chakraborti2021blast}  and two dimensions~\cite{singh2023blast} for the case of an explosion in a homogeneous gas.

In this paper, we generalize these results to the case of an explosion in an inhomogeneous gas, where the initial density varies as a power law $\rho(r)=\rho_0 r^{-\beta}$. We generalize the exact solution of the Euler equation to $d$-dimensions. By examining the asymptotic behavior near the shock center, we identify  a critical $\beta$ for which we conjecture that Euler equation should give a complete description of the problem both at the shock center as well as the shock front. For other $\beta$, we show that there is a crossover behavior near the shock center. We generalize the results of Ref.~\cite{ganapa2021blast,chakraborti2021blast,singh2023blast} for $\beta=0$ to arbitrary $\beta$ to derive a crossover length scale as well as a core scaling which describes the data near the shock center. These results are verified using event driven molecular dynamics (EDMD) simulations in one dimension and numerical integration of the Navier-Stokes equation in one dimension.

The remainder of the paper is organized as follows. In \sref{sec2-reviewEE}, we generalize the exact solution of Euler equation to $d$-dimensions, and find the asymptotic behavior of the thermodynamic quantities near the shock center. Based on these results, we derive a critical $\beta_c$ for which the solution satisfies $\nabla T=0$ at the center.  In \sref{sec3-modelEDMD} we define the model for the EDMD simulations, and show that the simulation data differs from the exact solution near the shock center for the different thermodynamic quantities. In \sref{sec4-NSE}, after describing the details of the  numerical solution of the Navier-Stokes equation, we show that including dissipation terms are able to reproduce the EDMD data for different $\beta$. The crossover near the shock center is studied in \sref{sec7-corescaling} by identifying a crossover length scale and the resultant scaling of the different thermodynamic functions. These are verified in EDMD simulations. In \sref{sec6-criticalTQ}, we show numerically that the data for critical $\beta$ are completely described by the Euler equation without any dissipation terms. We conclude with a summary and discussion in  \sref{sec8-summary}.

\section{\label{sec:problem} Problem definition}

At initial time $t=0$, we consider a gas whose density is inhomogeneous and varies from the center as
\be
\rho(\vec{r}, t=0) = \frac{\rho_0}{r^{\beta}},
\label{eq:initialdensity}
\ee
where $\beta$ is the exponent characterizing the power law variation. To ensure that the total mass in a bounded region is finite,  we require $\beta<d$, where $d$ is the dimension~\cite{sedov_book}. 

Energy $E_0$ is isotropically input at the center at time $t=0$. The "explosion" results in the formation of a spherically symmetric shock that propagates outwards in time with a shock front at $R(t)$ separating the disturbed gas from the ambient gas. The radius of shock front $R(t)$ can be obtained in terms of the parameters $E_0$, $\rho_0$, and time $t$ using dimensional analysis~\cite{landaubook,sedov_book,stanyukovich2016unsteady}:
\be
R(t) \propto \left(\frac{E_0}{\rho_0}\right)^{\alpha/2}t^{\alpha}, \label{frontradius}
\ee
where 
\be
\alpha=\frac{2}{2+d-\beta}.
\ee

We study the self-similar evolution of the shock using different methods: within Euler equation in $d$-dimensions, EDMD simulations of point sized binary gas in one dimension and Navier-Stokes equation in one dimension. We will be primarily interested in the following thermodynamic quantities: temperature $T(r,t)$, radial velocity $u(r,t)$, and density $\rho(r,t)$. In particular, we will analyze the scaling of these quantities near the shock front as well as the core region near the center of the shock.

\section{\label{sec2-reviewEE} Exact solution of the Euler equation in $d$-dimensions}

\subsection{Euler equation}

In this section, we generalize the known exact solution of the Euler equation for an explosion in an ideal gas with a density gradient. This will also allow us to derive the behavior of the thermodynamic quantities near the shock center, which in turn will allow us to determine the crossover exponents for $r \to 0$ in the presence of dissipative terms.

The microscopic description of the gas, at position $\vec{r}$, and at time $t$, is given by the local fields of the density $\rho(\vec{r},t)$, velocity $\vec{u}(\vec{r},t)$, temperature $T(\vec{r},t)$, and the pressure $p(\vec{r},t)$. These thermodynamic quantities evolve in time based on the continuity equations of the mass, momentum, and the energy. In the scaling limit $r,t \to \infty$, keeping $r t^{-\alpha}$ constant [see Eq.~(\ref{frontradius})], the contributions of heat conduction and viscosity terms become negligible. The continuity equations, after ignoring these terms, result in the Euler equation. Using spherical symmetry, the Euler equation in radial coordinates in $d$-dimensions is given by~\cite{landaubook,barenblatt1996scaling,whitham2011linear,sedov_book,dokuchaev2002self},
\begin{align}
&\partial_t \rho + \partial_r (\rho u) +\frac{(d-1)\rho u}{r} =0,\label{masseq}\\
&\partial_t u + u \partial_r u + \frac{1}{\rho} \partial_r p =0, \label{momentumeq}\\
&\partial_t \left( \frac{p}{\rho^\gamma}\right) + u \partial_r \left( \frac{p}{\rho^\gamma}\right)=0, \label{energyeq}
\end{align}
where $\gamma=1+2/d$ is the adiabatic index of a mono-atomic gas. If  local thermal equilibrium of the gas is considered, then the local pressure of the gas can be obtained from the equation of the state of the gas,
\begin{align}
p=k_B \rho T,  \label{idealEOS}
\end{align}
where $k_B$, the Boltzmann constant, is set equal to one in the remainder of the paper.

We introduce non-dimensional distance $\xi$ , density $\widetilde{R}$, velocity $\widetilde{u}$, and  temperature $\widetilde{T}$~\cite{sedov_book,dokuchaev2002self}: 
\begin{align}
&\rho(r,t)=\rho_0 r^{-\beta} \widetilde{R}(\xi), \label{rescadens} \\
&u(r,t)= \frac{r}{t}\widetilde{u}(\xi), \label{rescavel}\\
&T(r,t)= \frac{r^2}{t^2}\widetilde{T}(\xi),\label{rescatemp}\\
&\xi=r \left(\frac{E_0}{\rho_0}\right)^{-\alpha/2}t^{-\alpha}. \label{rescadist} 
\end{align}
The position of the shock front will be denoted by $\xi_f$. The Euler equation~(\ref{masseq})--(\ref{energyeq}), on non-dimensionalizing reduce to ordinary differential equations:
\begin{align}
&(\widetilde{u}-\alpha) \frac{d \log \widetilde{R}}{d \log \xi} + \frac{d \widetilde{u}}{d \log \xi} + (d-\beta)\widetilde{u}=0,\label{idealmass}\\
&(\widetilde{u}-\alpha) \frac{d \widetilde{u}}{d \log \xi} + \frac{d \widetilde{T}}{d \log \xi}+ \widetilde{T} \frac{d \log \widetilde{R}}{d \log \xi} +(2-\beta) \widetilde{T} + \widetilde{u}^2 -\widetilde{u} =0,\label{idealmomentum}\\
&(\widetilde{u}-\alpha)\frac{d}{d \log \xi} \log\left( \frac{\widetilde{T}}{{\widetilde{R}}^{\gamma-1}}\right) + [2+\beta(\gamma-1)]\widetilde{u} -2=0.\label{idealenergy}
\end{align}

Across the shock front, the thermodynamic quantities are discontinuous and obey the Rankine-Hugoniot boundary conditions, which are~\cite{landaubook,whitham2011linear,sedov_book},
\begin{align}
&\rho_1=\left(\frac{\gamma+1}{\gamma-1}\right)\rho_0,\label{RHdens}\\
&u_1=\frac{2}{\gamma+1}U,\label{RHvel}\\
&p_1=\rho_0 U u_1, \label{RHpress}
\end{align}
where the subscript $1$ signifies values just behind the shock front, and $U=\dot{R}$ is the speed of the shock front. In terms of the dimensionless quantities, the Rankine-Hugoniot boundary conditions~(\ref{RHdens})--(\ref{RHpress}) reduce to
\begin{align}
&\widetilde{R}(\xi_f)=\frac{\gamma+1}{\gamma-1},\label{idealRHmass}\\
&\widetilde{u}(\xi_f)=\frac{2\alpha}{\gamma+1},\label{idealRHpress}\\
&\widetilde{T}(\xi_f)= \frac{2\alpha^2(\gamma-1)}{(\gamma+1)^2}.\label{idealRHenergy}
\end{align}

We need one more equation for determining $\xi_f$. This is provided by the energy conservation, ie, the total energy is $E_0$. This constraint reduces to 
\begin{align}
S_d\int_0^{\xi_f}\left(\frac{1}{2}\widetilde{R}\widetilde{u}^2+\frac{\widetilde{R}\widetilde{T}}{\gamma-1}\right) \xi^{d+1-\beta} d\xi=1, \label{idealbeta}
\end{align}
where $S_d=\frac{2\pi^{d/2}}{\Gamma(d/2)}$ is the surface area of $d$-dimensional sphere of unit radius. $\Gamma$ is the Gamma function.

It has been shown~\cite{landaubook,sedov_book} that the solution curve of Euler equation (\ref{idealmass})--(\ref{idealenergy}) in the $\widetilde{T}$-$\widetilde{u}$ plane, that passes through the Rankine-Hugoniot boundary conditions~(\ref{RHdens}-\ref{RHpress}), should satisfy
\begin{align}
\widetilde{T} = \frac{\widetilde{u}^2(\alpha-\widetilde{u})(\gamma-1)}{2(\gamma \widetilde{u}-\alpha))},\label{idealenergyintegral}
\end{align}

\subsection{The exact solution}

The analytical solution of Eqs.~(\ref{idealmass})--(\ref{idealmomentum}) with the help of integral curve (\ref{idealenergyintegral}) using Rankine-Hugoniot boundary Eqs.~(\ref{idealRHmass})--(\ref{idealRHenergy}) is given by 
\begin{align}
&\frac{\xi (\widetilde{u})}{\xi_f}= \alpha^\alpha 2^\frac{a_2}{2a_1}  \frac{ \left(g_3(\widetilde{u})\right)^\frac{\gamma-1}{a_0}}{\left((\gamma +1) \widetilde{u}\right)^{\alpha }} \left(\frac{g_1(\widetilde{u})}{\alpha b_1}\right)^\frac{a_4}{2a_1} e^{\frac{ a_3 \left(\tanh ^{-1}\left(\frac{g_2(\widetilde{u})}{\sqrt{A_1}}\right)-\tanh ^{-1}\left(\frac{g_2(\widetilde{u}(\xi_f))}{\sqrt{A_1}}\right)\right)}{a_1 \sqrt{A_1}}}, \label{exactxi}\\ 
&\widetilde{R}(\widetilde{u})= \alpha \frac{\left(g_3(\widetilde{u})\right)^\frac{d-\beta}{a_0}}{\alpha-\widetilde{u}} \left(\frac{g_1(\widetilde{u})}{2 \alpha b_1}\right)^\frac{b_2}{2a_1} e^{\frac{ b_3\left(\tanh ^{-1}\left(\frac{g_2(\widetilde{u})}{\sqrt{A_1}}\right)-\tanh ^{-1}\left(\frac{g_2(\widetilde{u}(\xi_f))}{\sqrt{A_1}}\right)\right)}{a_1 \sqrt{A_1}}}, \label{exactg}
\end{align}
where
\begin{align}
\begin{split}
&A_1=\alpha ^2 (\beta  (\gamma -1)-2 \gamma +4)^2+4 \gamma ^2+4 \alpha  \left((\beta -2) \gamma ^2-(\beta -4) \gamma -2 (\gamma -1) d-4\right),\\
&a_0=d-(\beta -2) \gamma-2,\\
&a_1= (d (\gamma -1)+2) (d-(\beta -2) \gamma -2),\\
&a_2=\alpha  (\gamma -1) d^2-\left(\alpha  \left((\beta -2) \gamma ^2-(\beta -4) \gamma -4\right)-(\gamma -1)\gamma -2\right) d \nonumber\\
&\qquad - (\beta-2)(\gamma +1) \gamma -2 \alpha  ((\beta -2) \gamma +2)-4,\\
&a_3= a_1 ((\beta-2)\gamma -\beta +4)\alpha ^2 + \left[ 2 \left(\gamma ^2-5 \gamma +4\right) d^2 -\beta  \left(\gamma ^3-8 \gamma ^2+5 \gamma +2\right) d \right. \nonumber \\
&\qquad +2 \left(\gamma ^3-9 \gamma ^2+16 \gamma -12\right) d + \beta ^2 \left(\gamma -\gamma ^3\right) +\beta  \left(4 \gamma ^3-6 \gamma ^2+6 \gamma +4\right) \nonumber \\
&\qquad  \left. -4 \left(\gamma ^3-3 \gamma ^2+6 \gamma -4\right) \right] \alpha - 2 \gamma ^2 (\gamma  \beta +\beta +d (\gamma -3)-2 \gamma +2), \\
&a_4=\alpha  (\gamma -1) d^2-\left(\alpha  \left((\beta -2) \gamma ^2-(\beta -4) \gamma -4\right)+(\gamma-1)\gamma +2\right) d \nonumber\\
&\qquad + (\beta-2)(\gamma +1) \gamma -2 \alpha  ((\beta -2) \gamma +2)+4,\\
&b_1=-\frac{(\gamma -1) (\alpha  ((\beta-2)  \gamma +\beta -2 d )+\gamma +1)}{(\gamma +1)^2},\\
&b_2=\beta  \left((\beta-2)\gamma^2+\beta \gamma +4\right)+2 d^2-d \left((\beta-2)\gamma^2 + \left(\gamma +2\right)\beta +4\right),\\
&b_3=(d-\beta ) ((\beta-2) \gamma +\beta-2d) \left(\alpha  \left((\beta -2) \gamma ^2-(\beta -4) \gamma -2 (\gamma -1) d-4\right)+2 \gamma ^2\right),\\
&g_1(\widetilde{u})=\widetilde{u}^2 ((\gamma -1) d+2)-\widetilde{u} (\alpha  ((\beta-2)\gamma -\beta +4)+2 \gamma )+2\alpha,\\
&g_2(\widetilde{u})=2 \widetilde{u} ((\gamma -1) d+2) - \alpha ((\beta-2) \gamma -\beta +4)-2 \gamma ,\\
&g_3(\widetilde{u})=\frac{(\gamma +1)(\gamma \widetilde{u}-\alpha)}{\alpha(\gamma-1)}.
\end{split}
\end{align}

The above constants depend on $\beta$. The constants $a_0$ and $b_1$ become zero for certain values of $\beta$ resulting in the solution becoming singular. These values of $\beta$ are tabulated in \tref{table1}. For these special cases, the exact solution has been described in $d=3$~\cite{sedov_book}. In this paper, we will consider only the generic non-singular case.
\begin{table}
\begin{center}
\caption{\label{table1} 
The values of $\beta$ in different dimensions for which the constants $a_0$ or $b_1$ equals zero, making the analytical solution in Eqs.~(\ref{exactxi})--(\ref{exactg}) singular.} 
\begin{tabular}{cccc}
  &$\beta(d=1)$  & $\beta(d=2)$  & $\beta(d=3)$\\
  \hline
 $a_0$ & ${5}/{3}$ & $2$ & ${13}/{5}$\\
 $b_1$ & $1$ & ${4}/{3}$ & $2$\\
 \hline
\end{tabular}
\end{center}
\end{table}

\subsection{Asymptotic behavior for $\xi\to 0$}

According to the solution \eref{exactxi}, when $\xi\to 0$ then $\widetilde{u} \to \frac{\alpha}{\gamma}$, and the range being $\widetilde{u}\in \left[\alpha/\gamma, 2\alpha/(\gamma+1)\right]$. When $\xi \to 0$, the exact solutions of $\widetilde{u}$, $\widetilde{R}$, and $\widetilde{T}$ lead to the following asymptotic behavior:
\begin{align}
&\widetilde{u} -\frac{\alpha}{\gamma} \to \xi^{\frac{d-2+(2-\beta)\gamma}{\gamma-1}}, \label{asy_V}\\
&\widetilde{R} \to \xi^\frac{d-\beta}{\gamma-1}, \qquad \xi \to 0,\label{asy_G}\\
&\widetilde{T} \to \xi^{-\frac{d-2+(2-\beta)\gamma}{\gamma-1}}. \label{asy_Z} 
\end{align}
The behavior near the shock center is dependent on  the inhomogeneity parameter $\beta$, unlike on the driving parameter in driven shock~\cite{kumar2024shock}. For $\beta =0$ all of these exponents become equal to those for single impact~\cite{joy2021shock3d,joy2021shock,singh2023blast}.

It is convenient to define a different non-dimensional velocity $\widetilde{V}$ instead of $\widetilde{u}$  to avoid numerical difficulties in measuring $\widetilde{u}$ near the center. We define $\widetilde{V}$ as
\begin{align}
&\widetilde{V}= \xi \widetilde{u}.\label{tvnsvelxi}
\end{align}
Near the shock center, $\widetilde{V}$ follows the following asymptotic behavior,
\begin{align}
&\widetilde{V} -\frac{\alpha \xi}{\gamma} \to \xi^{\frac{d-3+(3-\beta)\gamma}{\gamma-1}} \label{asy_Vxi}
\end{align}

\begin{figure}
\includegraphics[scale=0.265]{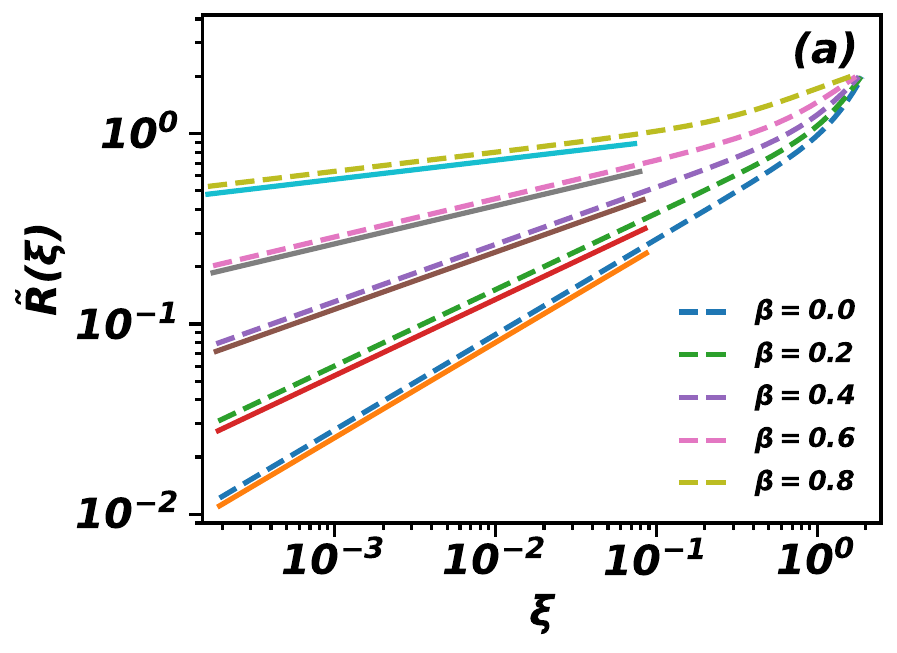}
\includegraphics[scale=0.265]{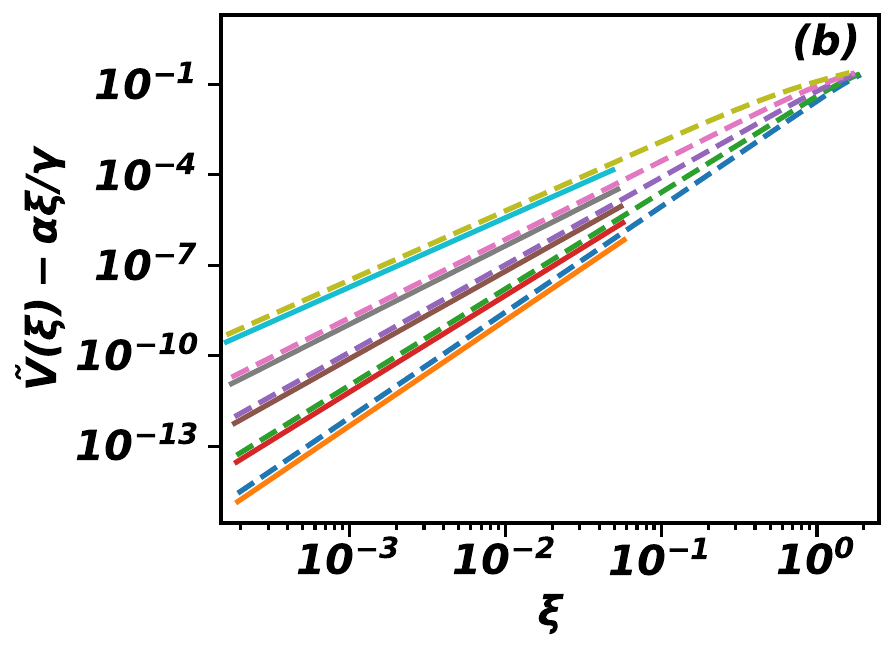}
\includegraphics[scale=0.265]{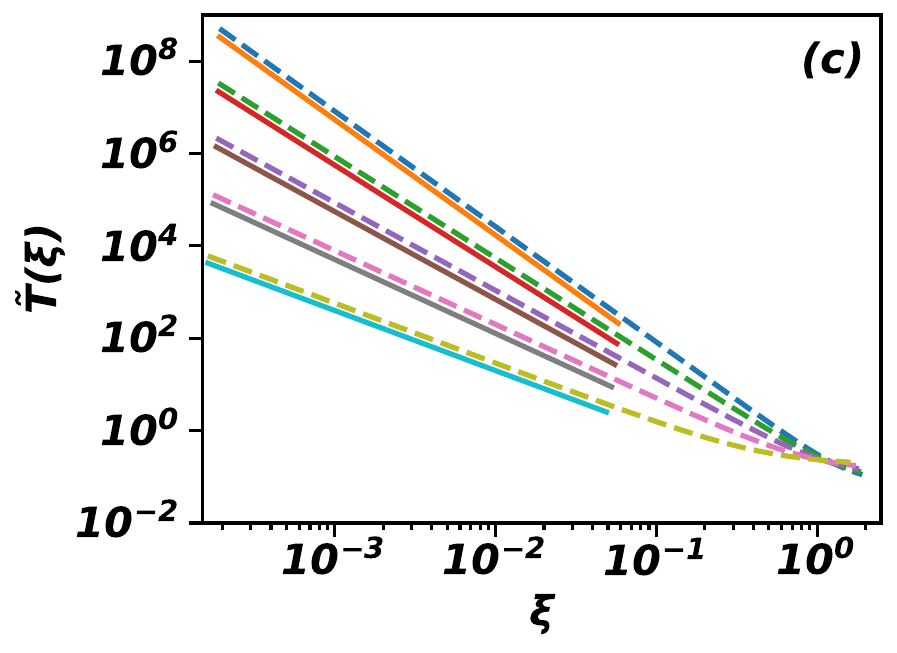}
\caption{(a) density $\widetilde{R}$, (b) velocity $\widetilde{V}$, and (c) temperature $\widetilde{T}$ obtained from the exact solution [see Eqs.~(\ref{idealenergyintegral})--(\ref{exactg})] of the Euler equation for five different values of $\beta=0.0$, $0.2$, $0.4$, $0.6$, $0.8$. The colored solid lines represent the fitting of power law behavior near the shock center (see Eqs.~(\ref{asy_G}-\ref{asy_Z}) and \eref{asy_Vxi}) for various $\beta$. All the curves are for $d=1$.}
\label{power_laws}
\end{figure}

To confirm the correctness of the asymptotic behavior near the shock center, we compare the power law behavior with the exact solution in  \fref{power_laws} for  five different values of $\beta$. The asymptotic behavior compare well with the exact result for  $\widetilde{R}$ [\fref{power_laws}(a)], $\widetilde{V}$ [\fref{power_laws}(b)], and $\widetilde{T}$ [\fref{power_laws}(c)].

\subsection{Critical $\beta$ \label{sec:critical}}

In the exact solution, the power law exponent of temperature depends on $\beta$ [see \eref{asy_Z}], and temperature diverges when $\xi \to 0$. We expect that this behavior cannot be correct and that in the presence of heat conduction, the radial derivative of temperature should be zero. We identify the value of $\beta$ for which the solution to the Euler equation satisfies $\nabla T=0$ to be a critical beta $\beta_c$. We conjecture that for $\beta_c$, the Euler equation should give a complete description of the problem.  For $\nabla T \to 0$, we require that $\widetilde{T} \to \xi^{-2}$. From Eq.~(\ref{asy_Z}), we immediately obtain the value of $\beta_c$ in $d$-dimensions to be
\begin{align}
\beta_c=\frac{d}{\gamma}.
\end{align}
We will verify this conjecture with simulations in later sections.

\section{\label{sec3-modelEDMD} Molecular Dynamics Simulations }

In this section, we check the asymptotic behaviors Eqs.~(\ref{asy_G}-\ref{asy_Z}) and \eref{asy_Vxi} using EDMD simulation of hard point particles in one-dimension with adjacent particles having different masses $m_1$ and $m_2$. 

\subsection{Model}

Consider a system of $N$ point particles labeled $i=1$, $2$, $3$, $\ldots$, $N$ sorted according to their positions $x_i$. Let their velocities be denoted by  $u_i$. All the particles with odd $i$ have mass $m_1$ and even $i$ have  mass $m_2$. Particles do ballistic motion  until they undergo  energy and momentum conserving elastic collisions. Since the particles are on a line, the particle $i$ can only collide with the particles $i-1$ or $i+1$, which conserves the ordering of the particles. If $u_i$, and $u_j$ are the pre-collision velocities of the particles $i$, and $j$ then their post-collision velocities $u_i'$, and $u_j'$, respectively, are given by
\begin{align}
u_i'=\frac{m_i u_i +  m_j u_j + m_j(u_j-u_i)}{m_i+m_j},\\
u_j'=\frac{m_j u_j +  m_i u_i + m_i(u_i-u_j)}{m_i+m_j}.
\end{align}

We consider the region $0\le x \le L$ centered about with $x=L/2$. To implement the initial density as given in Eq.~(\ref{eq:initialdensity}), we divide the region in equally spaced bins with bin-size $\Delta r$. The number of particles, $N_r$, in a bin at distance $r$ from $L/2$  is given by
\begin{align}
N_r= \left\lceil\frac{N}{2(L/2)^{1-\beta}}\left[(r+\Delta r)^{1-\beta}-r^{1-\beta}\right]\right\rceil, \label{numparticlesedmd}
\end{align}
where $\lceil \hdots \rceil$ is the least integer function. This 
gives $\rho_0 = \frac{N}{4}\frac{(1-\beta)(m_1+m_2)}{(L/2)^{(1-\beta)}}$. The $N_r$ particles are distributed at random  in the respective bins in $[0,L/2]$. Symmetry about $L/2$ is ensured by filling the bins in $[L/2, L]$ based on the positions of particles in bins in $[0,L/2]$.

All the particles are initially at rest. An initial energy $E_0$ is input by giving  to $N_c$ number of particles around the center a non-zero velocity that depends on their position. The first $N_c/2$ particles are given velocities as
\begin{align}
u_{N/2+1+i}=u_0 e^\frac{-(x_{N/2+1+i}-L/2)^2}{2\sigma^2},
\end{align} 
and the remaining $N_c/2$ particles have velocity $u_{(N-N_c)/2+i}= -u_{(N+N_c)/2+1-i}$, where $i \in [0,N_c/2]$, and $u_0$, $\sigma$ are the positive constants. The velocities are rescaled  to ensure that the system has total energy $E_0$, and zero linear momentum.

In the EDMD simulations for different $\beta$, we take $N=50000$, $L=10000$, and  $E_0=24$, $N_c=32$, $m_1=1$, $m_2=2$. The simulation time is chosen such that the shock front does not reach the boundaries.

\subsection{Behavior of thermodynamic quantities}

\begin{figure}
\begin{center}
\includegraphics[scale=0.35]{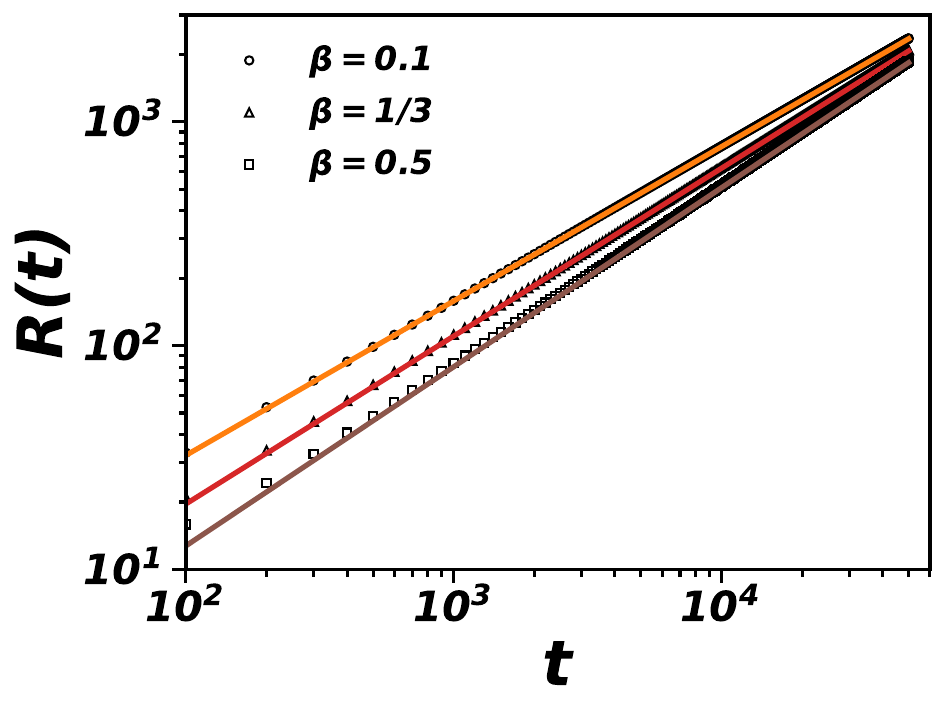}
\caption{ The power law variation of  the radius of shock front, $R(t)$, with time $t$. Symbols represent the EDMD data for three different values of $\beta=0.1$, $1/3$, $0.5$. The solid lines represent the fitting of power law behavior $R(t)\sim t^{\frac{2}{3-\beta}}$.}
\label{densityradiuspower_laws}
\end{center}
\end{figure}

To benchmark the EDMD simulations, we reproduce the scaling of the shock front, $R(t)$ with time, as given in Eq.~(\ref{frontradius}). Our simulations reproduce the scaling exponent for different $\beta$ accurately as can be seen from \fref{densityradiuspower_laws}.

To measure  $\rho(x,t)$, $u(x,t)$, $T(x,t)$, we divide the system size into bins with bin size $\Delta'$, and measure these quantities in a bin at spatial position $x\in [0,L]$ as
\begin{align}
&\rho(x,t)=\left\langle\frac{\Sigma m_i \delta (x_i,x)}{\Delta'}\right\rangle,\label{densityedmd}\\
&u(x,t)=\left\langle\frac{\Sigma m_i u_i\delta (x_i,x)}{\Sigma m_i \delta (x_i,x)}\right\rangle,\label{velocityedmd}\\
&T(x,t)=\left\langle\frac{\Sigma m_i u_i^2 \delta (x_i,x)}{\Sigma m_i \delta (x_i,x)}\right\rangle-\left(\left\langle\frac{\Sigma m_i u_i\delta (x_i,x)}{\Sigma m_i \delta (x_i,x)}\right\rangle\right)^2,\label{tempedmd}
\end{align}
where $\langle\hdots\rangle$ denotes averaging over different initial configuration, and $\delta$ is a step function defined as
\begin{align}
& \delta(x_i,x) = \left\{
\begin{array}{ll}
      1 &  |x-x_i| \le \Delta'/2  \\
      0 &  |x-x_i| > \Delta'/2 \\
\end{array} \right.
\end{align}
We average over $10^3$ different initial configurations for each $\beta$. We  present data for mostly $\beta=0.1$, $0.5$. We note that in one dimension $\beta_c=1/3$, and these two values are chosen as representative values above and below $\beta_c$. 

\begin{figure}
\centering
\includegraphics[scale=0.2]{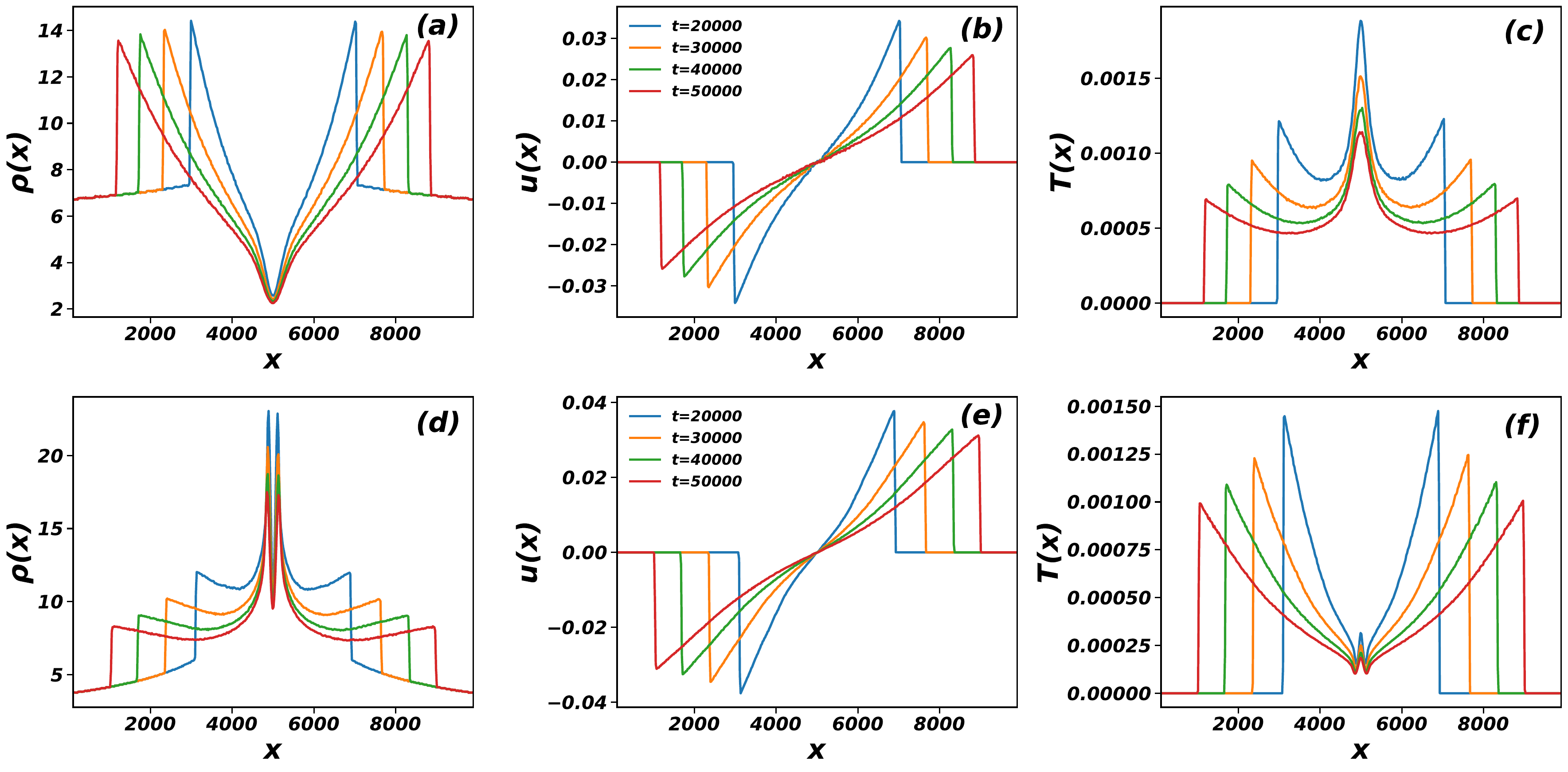}
\caption{ The variation of  (a) density $\rho(x,t)$, (b) velocity $u(x, t)$, and (c) temperature $T(x, t)$ with spatial position $x$ at four different times $t=20000$, $30000$, $40000$, $50000$. Figs.~(a)--(c) are for $\beta=0.1$, and Figs.~(d)--(f) are for $\beta=0.5$.}
\label{thermoprofile}
\end{figure}

The variation of $\rho$, $u$, $T$ with spatial position $x$ at four different times is shown in \fref{thermoprofile} for $\beta=0.1, 0.5$. The density $\rho(x)$ increases monotonically for $\beta=0.1$, while for $\beta=0.5$, there is a sharp increase in density at the shock center. Temperature also shows different behavior near the shock center. For $\beta=0.1$, temperature diverges near the shock center before being rounded off, while it is a minimum for $\beta=0.5$ at the shock center. The qualitative behavior of velocity $u(x)$ remains unchanged for all $\beta$.

\subsection{Comparison between the exact solution of the Euler equation  and simulations}

\begin{figure}
\centering
\includegraphics[scale=0.18]{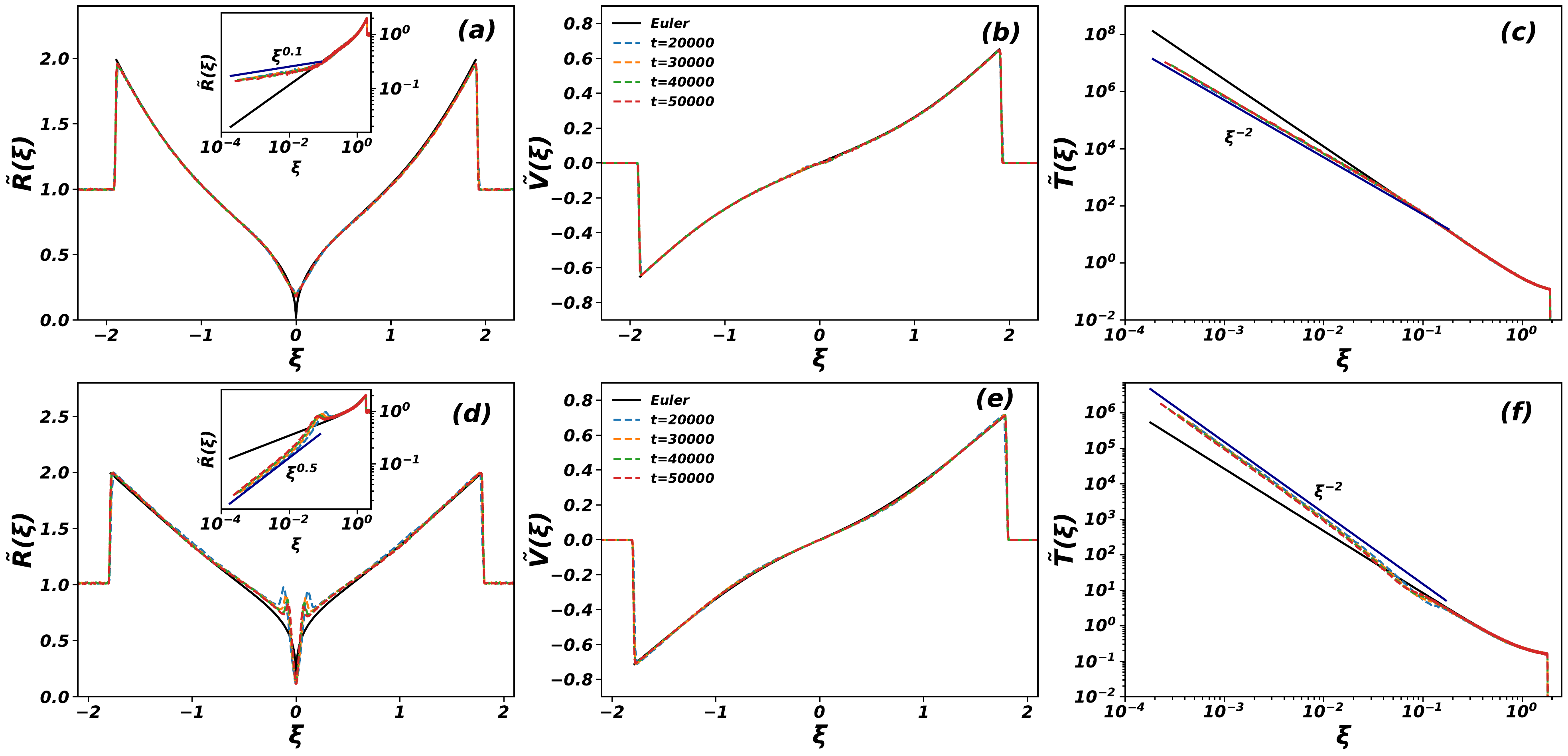}
\caption{ The comparison of $\widetilde{R}(\xi)$, $\widetilde{V}(\xi)$, and $\widetilde{T}(\xi)$ obtained from the EDMD simulations with the exact solution for two different values of $\beta=0.1$, $0.5$. Figs.~(a)-(c) represent the plot for $\beta=0.1$, and the Figs.~(d)-(f) represent the plots for $\beta=0.5$. The broken colored lines represent the EDMD results at four different times, and solid black lines represent the exact solution. Insets show the same data on the logarithmic scale. Solid dark blue lines represent the power law fitting of the data of density and temperature profiles obtained from the EDMD.}
\label{rescaledthermoprofile}
\end{figure}

In \fref{rescaledthermoprofile}, we compare the non-dimensionalized thermodynamic quantities $\widetilde{R}(\xi)$, $\widetilde{V}(\xi)$, and $\widetilde{T}(\xi)$ obtained from the EDMD simulations and the exact solution of Euler equation for two different values of $\beta=0.1$, $0.5$. First, we note that the data for the different times, away from the shock center,  collapse onto one curve,  validating the correctness of the scaling Eqs.(\ref{rescadens})--(\ref{rescadist}) for both values of $\beta$. The data away from the shock center match perfectly with the exact solution of the Euler equation.

The data near the shock center  do not seem to collapse (see the inset plots in \fref{rescaledthermoprofile}), which indicates that the system follows a different scaling near center. Near the shock center, the EDMD results show that the density $\widetilde{R}(\xi)$ increases as $\widetilde{R}\sim \xi^\beta$. Also, independent of $\beta$, temperature varies as $\widetilde{T}\sim \xi^{-2}$, or equivalently $\nabla T=0$. These power law exponents differ from the exponents given by the exact solution, $\widetilde{R}\sim \xi^{(1-\beta)/2}$, $\widetilde{T}\sim \xi^{-(5-3\beta)/2}$ , showing the discrepancy between the exact solution and the simulation results.

\section{\label{sec4-NSE}Navier-Stokes Equation}

In this section, we describe the numerical solution of Navier-Stokes equation, and its comparison with the EDMD results and the exact solution.

\subsection{Numerical details}

Taking into account the dissipation terms in the Euler equation, we obtain the  Navier-Stokes equation. The different continuity equations now reduce to~\cite{landaubook,warsi2005fluid},
\begin{align}
&\partial_t\rho +\partial_x(\rho u) = 0,\label{NSEmass}\\
&\partial_t(\rho u) +\partial_x \left(\rho u^2+p\right)  = \partial_x \left( \zeta \partial_x u \right),\label{NSEmom}\\
&\partial_t\Big(\frac{1}{2}\rho u^2 +  \frac{1}{2}\rho T\Big) +\partial_x\Big(\Big[\frac{1}{2}\rho u^2 +  \frac{1}{2}\rho T + p \Big]u\Big) = \partial_x\left( u \zeta \partial_x u + \lambda \partial_x T \right),\label{NSEene}
\end{align}
where $\zeta$ and $\lambda$ are the bulk viscosity, and the heat conduction of the system, respectively. According to the Green-Kubo relations, both of these quantities depend on temperature as $T^{1/2}$ in one dimension, while a recent study shows the heat conduction depend on density also as $\lambda \sim \rho^{1/3}$~\cite{hurtado2016violation}. Accordingly, we take these quantities as
\begin{align}
&\lambda= C_1\rho^{1/3}T^{1/2},\\
&\zeta= C_2 T^{1/2}.
\end{align}

We solve the Navier-Stokes equation~(\ref{NSEmass})--(\ref{NSEene}) in the region $-L/2\le x \le L/2$, numerically using MacCormack method~\cite{maccormack1982numerical}. MacCormack method provides an accuracy up to second order both in time discretization $\Delta t$, and space discretization $\Delta x$. We take the initial condition at time $t=0$ as : gradually decreasing density $\rho(x)=\rho_0 |x|^{-\beta}$ as in \eref{eq:initialdensity}, zero velocity everywhere, and the initial temperature as
\begin{align}
T(x,0)=T_0 e^\frac{-x^2}{2\sigma^2},
\end{align} 
We choose $L$ such that that the shock does not reach to the edge of the system at its maximum integration time $\tau$. The values of the parameters used in the numerical integration are tabulated in \tref{numericalparameters}.
\begin{table} 
\begin{center}
\caption{\label{numericalparameters} 
The values of parameters used in the  numerical solution of the Navier-Stokes Eqs.~(\ref{NSEmass})--(\ref{NSEene}).} 
\begin{tabular}{llll}
Parameters  & Values  \\
 \hline
 $L$ & $10000$ & \\
 $\tau$ & $80000$ & \\
 $T_0$ & $0.01$ \\
 $\sigma$ & $1$ \\
 $\Delta r$ & $0.1$ \\
 $\Delta t$ & $0.001$ \\
 $C_1$ & $5$ \\
 $C_2$ & $20$ \\
 \hline
\end{tabular}
\end{center}
\end{table}
 
\subsection{\label{sec5-comparisonexactEDMDdns}Comparison between Euler equation, Simulation, and  Navier-Stokes equation}

\begin{figure}
\centering
\includegraphics[scale=0.18]{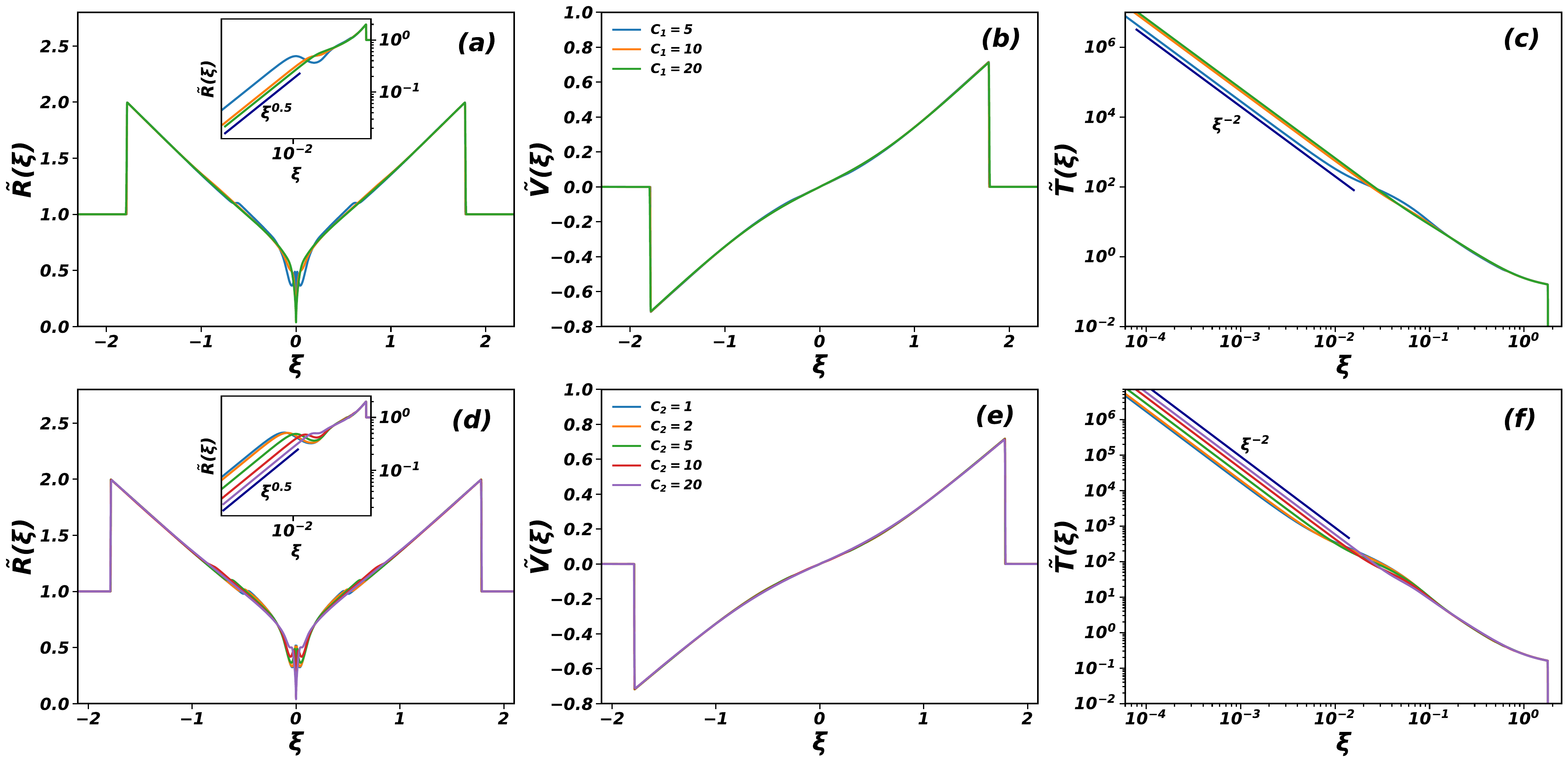}
\caption{The variation of $\widetilde{R}(\xi)$ (a,d),  $\widetilde{V}(\xi)$ (b,e), and $\widetilde{T}(\xi)$ (c,f) with $\xi$, obtained from the numerical solution of Navier-Stokes equation,  for $\beta=0.5$. Figs.~(a)--c) correspond to the  data for three different values of coefficient of heat conduction $C_1=5$, $10$, $20$ with $C_2=5$ fixed, and the Figs.~(d)--(f) correspond to the data for five different values of coefficient of bulk viscosity $C_2=1$, $2$, $5$, $10$, $20$ with $C_1=5$ fixed. The solid dark blue lines represent the power law fitting of the data of density and temperature profiles. The insets are the densities on the logarithmic scale.}
\label{eta_xeta}
\end{figure}

Since $C_1$ and $C_2$, the constants parameterizing heat conductivity and bulk visocity, are not known, we first perform a parametric study to understand their effect on the results. We fix both $\beta=0.5$ and time. We show the dependence of   $\widetilde{R}$, $\widetilde{V}$, and $\widetilde{T}$ on varying $C_1$ keeping $C_2$ fixed in  \fref{eta_xeta}(a)-(c) and on varying $C_2$ keeping $C_1$ fixed in  \fref{eta_xeta}(d)-(f). Introducing non-zero dissipation immediately changes the behavior near the shock center to the correct power laws as seen in the EDMD simulations. Changing $C_1$ and $C_2$ changes the results near the shock center quantitatively, keeping the exponent of the power law unchanged. The velocity is unaffected by dissipation. Changing the values of the coefficients does not affect the results near the shock front.  The change near the center indicates the heat conduction becomes important in the tiny region $0\le |x| \le X(t)$ (see \sref{corescaling}) and becomes negligible in the region $X(t) \le |x| \le R(t)$. 

\begin{figure}
\centering
\includegraphics[scale=0.18]{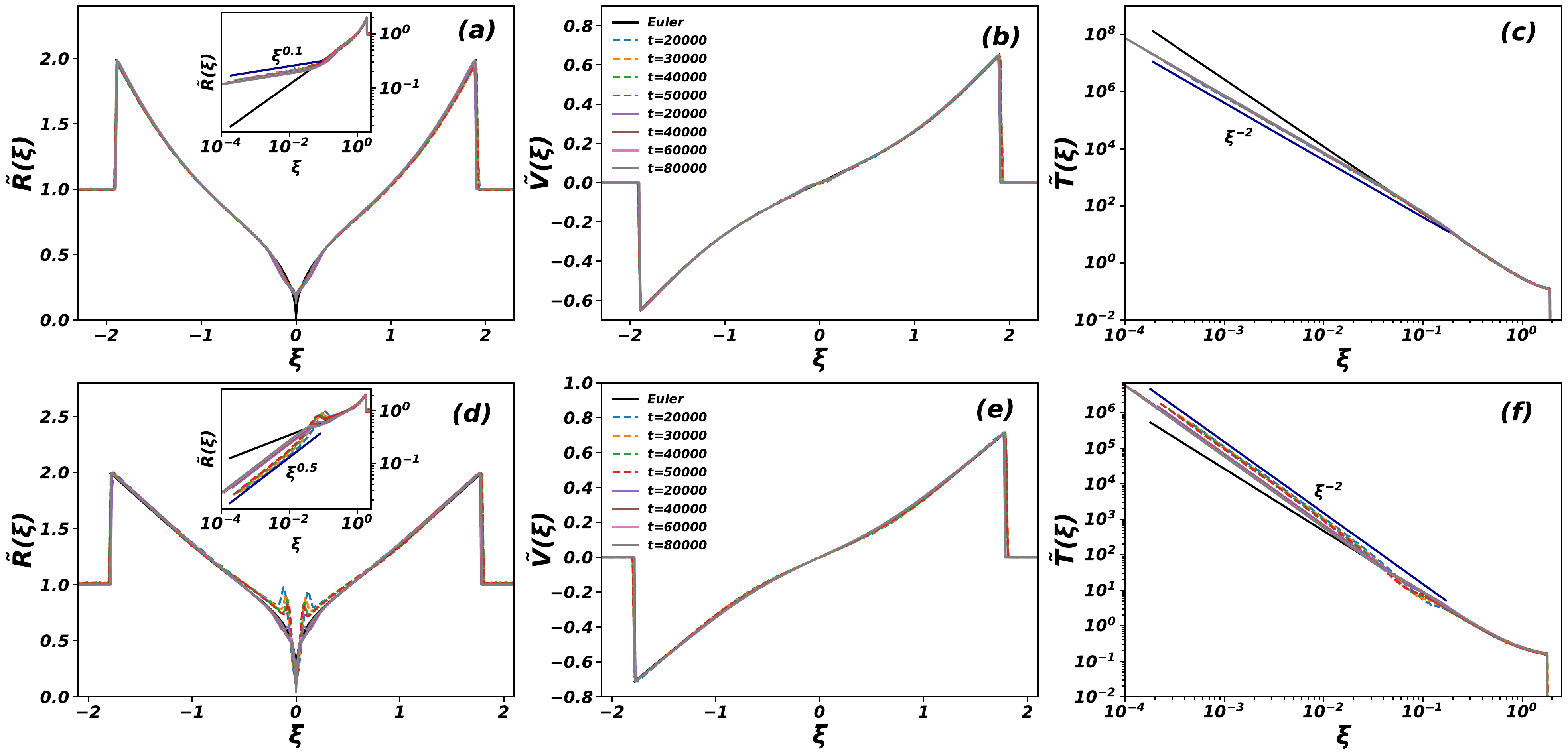}
\caption{The comparison of density $\widetilde{R}(\xi)$, velocity $\widetilde{V}(\xi)$, and temperature $\widetilde{T}(\xi)$ obtained from the EDMD results with the solution of Euler equation, and the numerical integration of Navier-Stokes equation. Figs.~(a)-(c) represent the plots for $\beta=0.1$, and the Figs.~(d)-(f) represent the plots for $\beta=0.5$. The dashed colored lines represent the EDMD results, solid black lines represent the exact solution of Euler equation, and the solid colored lines represent the numerical integration of Navier-Stokes equation . The insets are the density plots on logarithmic scale. Solid dark blue lines represent the power law fitting of the data of density and temperature. For the integration of the Navier-Stokes equation,  $\beta=0.1$ with $\rho_0=15.8$, and $\beta=0.5$ with $\rho_0=260$ and $C_1=5$, $C_2=20$.}
\label{rescaledthermoprofiledns}
\end{figure}

In \fref{rescaledthermoprofiledns}, we compare the  $\widetilde{R}$, $\widetilde{V}$, and $\widetilde{T}$ obtained from the exact solution of the Euler equation, EDMD, and the numerical solution of the Navier-Stokes equation for $\beta=0.1$, $0.5$, each for four different times. The Navier-Stokes data are for the choice $C_1=5$, $C_2=20$, which we found to be a good approximation to the solution. Near the shock center, the EDMD results and the Navier-Stokes results show the same power law behavior of density as $\widetilde{R}\sim \xi^\beta$ (inset plots), and the temperature as $\widetilde{T}\sim \xi^{-2}$ (Figs.(c), (f)), unlike the exact solution of the Euler equation. We find excellent agreement between the Navier-Stokes data and the EDMD data everywhere, which indicates that the Navier-Stokes equation is the correct description of the theory for the inhomogeneous gas also.

\section{\label{sec7-corescaling}Scaling near the shock center} \label{corescaling}
Due to inclusion of the heat conduction, an extra core scaling arises in the region $0\le |x| \le X(t)$, $X(t)$ being the size of the core~\cite{chakraborti2021blast}. We follow the procedure followed in Ref.~\cite{chakraborti2021blast} to find the crossover scaling for the homogeneous case, to find the core size $X(t)$ and the resultant scaling near the shock center for the inhomogeneous case.

In one dimension, near the shock center, the exact solution gives temperature and density as
\begin{align}
&T(x,t)\sim |x|^\frac{3 \beta -1}{2} t^\frac{1+\beta}{\beta-3}, \label{coretempeuler}\\
&\rho(x,t)\sim |x|^\frac{1-3\beta}{2} t^\frac{1-\beta}{\beta-3}. \label{coredenseuler}
\end{align} 
 In the region $0\le |x| \le X(t)$ the heat conduction dominates the Euler terms, while in the region $X(t) \le |x| \le R(t)$ Euler terms dominate the heat conduction term, showing the core size $r=X(t)$ be the location where both the terms become comparable. From the \eref{NSEene}, equating the Euler term with the heat conduction term i.e. $\frac{\rho T}{t} \sim \frac{\rho^{1/3} T^{3/2}}{|x|^2}$, and using the $T$ and $\rho$ obtained from the exact solution near the shock center and $|x|=X(t)$, we obtain the size of the core as
\begin{align}
&X(t)\sim t^\frac{2(19-13\beta)}{(3-\beta)(31-13\beta)}. \label{coresize}
\end{align}
\begin{figure}
\centering
\includegraphics[scale=0.18]{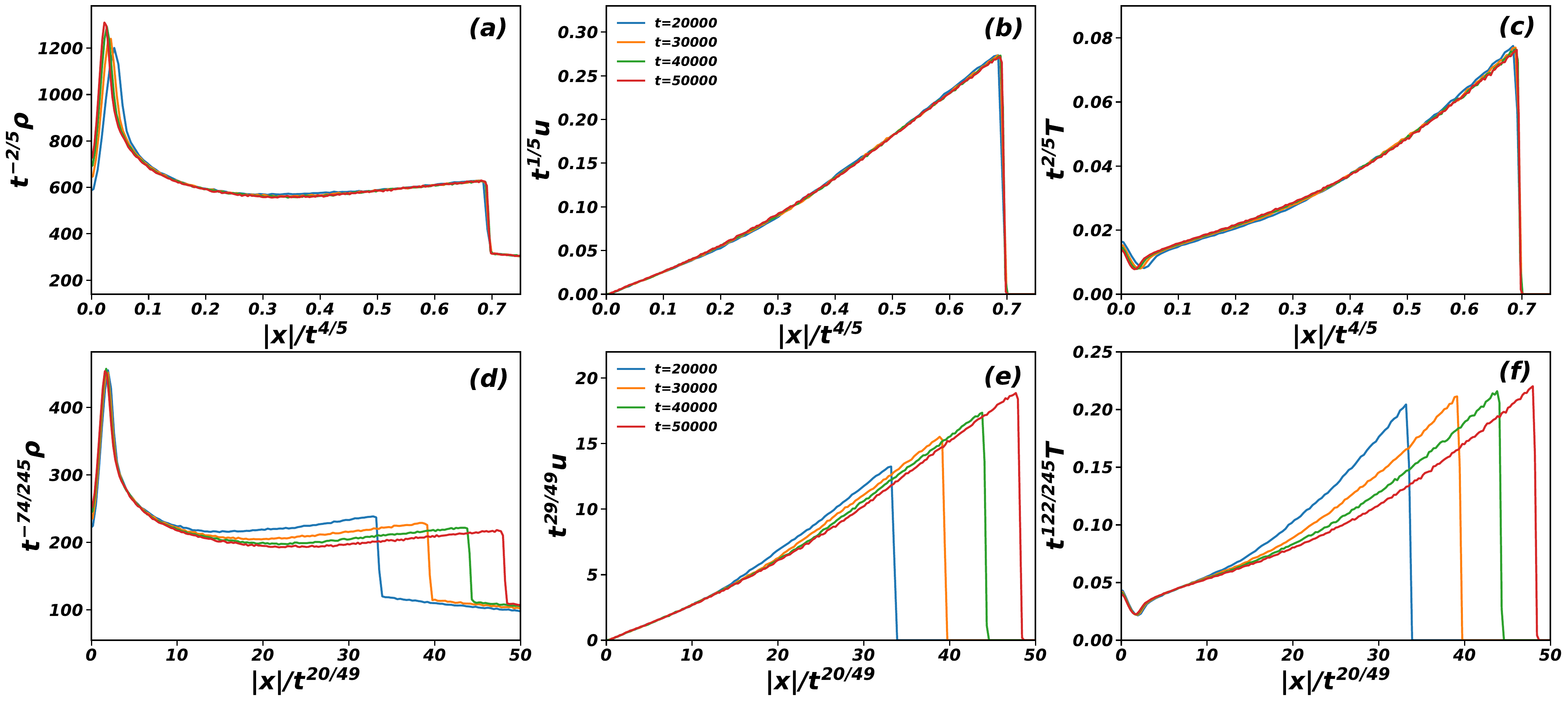}
\caption{The scalings of thermodynamic quantities of density $\rho$, velocity $u$, and temperature $T$ obtained from the EDMD results at four different times  for $\beta=0.5$ with $\rho_0=260$. Figs.~(a)-(c) represent the plots for front scaling Eqs.~(\ref{frontdens})--(\ref{fronttemp}), and the Figs.~(d)-(f) represent the plots for core scaling Eqs.~(\ref{coredens})--(\ref{coretemp}). Near the shock center, core scaling shows better collapse than front scaling.}
\label{corescalingplot}
\end{figure}

We now define a new rescaled distance $\xi' = x/X(t)$ near the shock center. With the help of Eqs.~(\ref{coretempeuler})--(\ref{coredenseuler}), and \eref{coresize}, we define the thermodynamic quantities near the shock center in terms of new scaling $\xi'$ as
\begin{align}
&\rho(x,t) \sim t^\frac{(1-3\beta)(19-13\beta)- (1-\beta)(31-13\beta)}{(3-\beta)(31-13\beta)} \widetilde{R}'(\xi'), \label{coredens}\\ 
&u(x,t) \sim t^\frac{2(19-13\beta)-(3-\beta)(31-13\beta)}{(3-\beta)(31-13\beta)} \widetilde{V}'(\xi')=\frac{X(t)}{t} \widetilde{V}'(\xi'), \label{corevel}\\
&T(x,t)\sim t^\frac{(3\beta-1)(19-13\beta)- (1+\beta)(31-13\beta)}{(3-\beta)(31-13\beta)} \widetilde{T}'(\xi'), \label{coretemp}
\end{align} 
while near the shock front, the system follows the scaling (see Eqs.~(\ref{rescadens}-\ref{rescadist})),
\begin{align}
&\rho(x,t) \sim t^\frac{2\beta}{3-\beta} \widetilde{R}^{''}(\xi), \label{frontdens}\\
&u(x,t) \sim t^\frac{\beta-1}{3-\beta} \widetilde{V}^{''}(\xi),\label{frontvel}\\
&T(x,t)\sim t^\frac{2(\beta-1)}{3-\beta} \widetilde{T}^{''}(\xi), \label{fronttemp}
\end{align}
where $\widetilde{R}'$, $\widetilde{V}'$, $\widetilde{T}'$, and $\widetilde{R}^{''}$, $\widetilde{V}^{''}$, $\widetilde{T}^{''}$ are the resultant non-dimensionalized thermodynamic quantities near the shock center, and near the shock front respectively.

We numerically confirm the scaling in \fref{corescalingplot}, where the scaling near the shock  front [Eqs.~(\ref{frontdens})--(\ref{fronttemp})] is shown in  \fref{corescalingplot}(a)-(c), while the  the core scaling [Eqs.~(\ref{coredens})--(\ref{coretemp})] is shown in \fref{corescalingplot}(d)-(f) for EDMD data for $\beta =0.5$. The data for different times show excellent data collapse in the appropriate regimes.

\section{\label{sec6-criticalTQ} Behavior for  critical $\beta_c$}

In Sec.~\ref{sec:critical}, we derived a critical $\beta_c=\gamma/d$. For this critical value of $\beta$, the solution of the Euler equation satisfies the boundary condition $\nabla T=0$ at the shock center. Hence, we expect that the Euler equation should provide a good description of the EDMD data both at the shock center as well as the shock front. 
\begin{figure}
\centering
\includegraphics[scale=0.18]{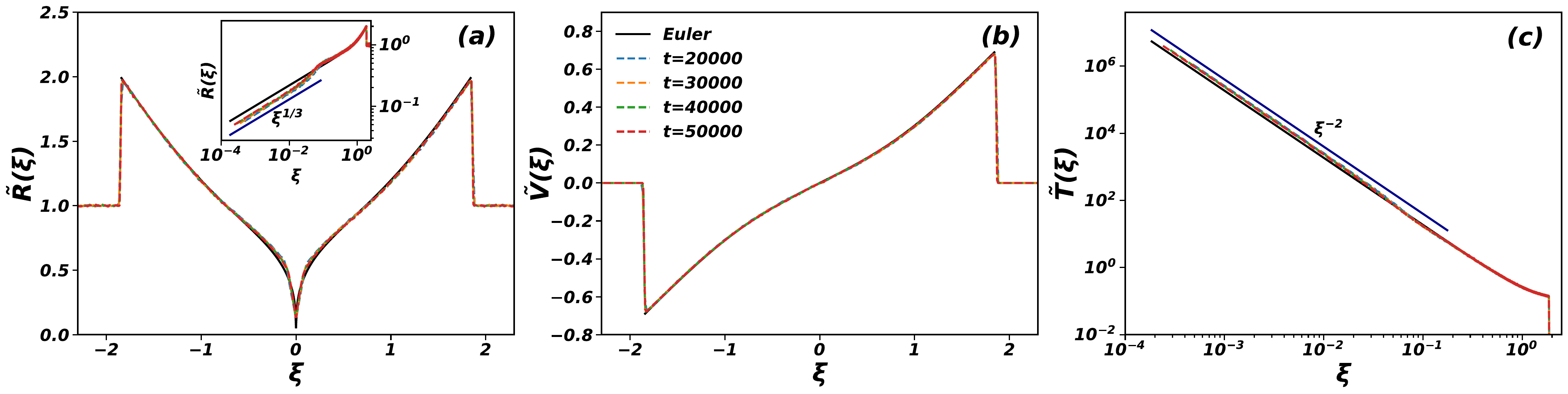}
\caption{For $\beta_c=1/3$, the EDMD data for (a) density $\widetilde{R}(\xi)$, (b) velocity $\widetilde{V}(\xi)$, and (c) temperature $\widetilde{T}(\xi)$ are compared with the exact result for the Euler equation. The dashed colored lines represent the EDMD data at four different times, and solid black lines represent the exact solution of Euler equation. The inset is the same data on logarithmic scale. Solid dark blue lines represent the power law fitting of the data of density and temperature.}
\label{rescaledthermocritical}
\end{figure}

We check numerically whether this is true for $\beta_c=1/3$ in $1$-dimension. $\widetilde{R}$, $\widetilde{V}$, and $\widetilde{T}$ obtained in EDMD simulations for four different times and $\beta=1/3$ are compared with the exact solution of the Euler equation in  \fref{rescaledthermocritical}. Excellent agreement is seen for all the thermodynamic quantities both at the shock front as well as the shock center.

\section{\label{sec8-summary}Summary and Discussion}

In summary, we studied the spatio-temporal evolution of density, velocity, and temperature following an explosion in an ideal gas with an initial  inhomogeneous density distribution $\rho(r)=\rho_0 r^{-\beta}$.  We generalized the exact solution of the Euler equation, consistent with the Rankine-Hugoniot boundary conditions, to $d$-dimensions. From the asymptotic behavior of the solution near the shock center, we argue that only for $\beta_c=\gamma/d$, should the Euler equation provide a full description of the problem. Using EDMD simulations in one dimension, we show that the Euler equation does not describe the data near the shock center. On the other hand the Navier-Stokes equation is able to over come this issue. The crossover length scale below which the dissipation terms are relevant are derived for arbitrary $\beta$. The core scaling for the data near the shock center are also derived for arbitrary $\beta$ and confirmed in EDMD simulations.

The results in this paper generalizes the known results for explosion in a homogeneous gas, a problem that is much better studied. By doing so, it is possible to pinpoint the exact reason why the Euler equation fails, even though it is the equation that satisfies the scaling limit. For the value of $\beta$ for which the solution satisfies $\nabla T=0$ at the shock center, the Euler equation is able to describe the data everywhere and not just near the shock front.

The propagation of shock has also been studied in the granular systems, whether generated by a single impact or by a continuous source. Examples include, the crater formation, due to the impact of high energy particles on a granular heap~\cite{grasselli2001crater}, or due to the vertical impact of a steel ball into the container of small glass beads~\cite{walsh2003morphology}, or due to the vertical impinges of gas jets on a granular bed~\cite{metzger2009craters}, and the shock propagation, due to the impact of a steel ball on a fast flowing granular layer~\cite{boudet2009blast}, or due to the sudden release of localized energy~\cite{jabeen2010universal,pathak2012shock}, or due to the continuous energy injection through continuous particle insertion~\cite{joy2017shock}, and the granular fingering and pattern formation due to the injection of viscous liquid into dry dense granular material~\cite{cheng2008towards,sandnes2007labyrinth,pinto2007granular,johnsen2006pattern,huang2012granular}. The hydrodynamic theory has been generalized to study the shock in granular gas, where energy is no longer a conserved quantity~\cite{barbier2015blast,barbier2016microscopic}.  It would be interesting to generalize these results to an inhomogeneous medium. 
%\bibliographystyle{spphys}
%\bibliography{Bibliography} 

\end{document}